\documentclass[%
aip,
amsmath,amssymb,
reprint
]{revtex4-1}
\usepackage{graphicx}% Include figure files
\usepackage{dcolumn}% Align table columns on decimal point
\usepackage{bm}% bold math
\usepackage{amsmath}
\usepackage{color}
\usepackage[normalem]{ulem}
\usepackage{balance}
\usepackage[T1]{fontenc}

\begin{document}

\preprint{AIP/123-QED}

\title{Understanding photoluminescence of coupled metallic nanostructures based on a coupling classic harmonic oscillator model}% Force line breaks with \\

\author{Yuqing Cheng}\affiliation{School of Mathematics and Physics, University of Science and Technology Beijing, Beijing 100083, China}%
\author{Mengtao Sun}\thanks{mengtaosun@ustb.edu.cn}
\affiliation{School of Mathematics and Physics, University of Science and Technology Beijing, Beijing 100083, China}%
\affiliation{Collaborative Innovation Center of Light Manipulations and Applications, Shandong Normal University, Jinan, 250358, China}

\date{\today}% It is always \today, today,

\begin{abstract}
Photoluminescence (PL) phenomenon from metallic nanostructures has been explained and understood by several point of views. One of them is based on the classic harmonic oscillator model, which describes PL of single mode. In this study, we continue to expand this classic model to a coupling case, which involves two oscillators that interact with each other together with the excitation electric field. The new generated modes due to the coupling are carefully analyzed, including their behaviors varying with the coupling coefficients in different cases. Furthermore, for practical purpose, PL spectra and white light scattering spectra of two individual metallic nanostuctures are calculated as examples employing the model to verify its validity. This work would give a deeper understanding on coupling PL phenomena and is helpful to relative applications.
\end{abstract}

\maketitle

%\tableofcontents

%\section{\label{sec:Introduction}Introduction}
%\textbf{\textit{Introduction}}.---
Photoluminescence (PL) phenomena from noble metals have been widely studied since the first report over 50 years ago \cite{PL1}. PL can be excited not only from bulk materials, but also from thin films and nanostructures \cite{PL2,PL3,PL4,PL5,PL6}. Particularly, the localized surface plasmon resonance (LSPR) effect enhances the emissions in the case of metallic nanostructures, thus resulting in numerous applications such as optical recording \cite{app1,app2}, biosensing \cite{app3,app4}, orientation probes \cite{app5,app6}, local temperature detection \cite{app9,app7,app8}.

The origin of PL has been discussed in plenty of studies, with different explanations such as interband transitions enhanced by LSPR \cite{mech1}, microscopic explanation for enhanced PL from gold nanoparticles \cite{mech2}, classic oscillator model assisted with electron distributions for single mode emission \cite{PL0}, and non-equilibrium electron dynamics affecting PL of metallic nanostructures \cite{mech3}. Nevertheless, the coupling PL phenomena are seldom investigated in theory. For example, Prodan E. \textit{et al}. present a molecular orbital theory to describe the coupling plasmon modes introduced by the metallic nanostructures of arbitrary shape \cite{cpm1}. Jain P. K. \textit{et al}. provide a semiempirical ``plasmon ruler equation'' based on discrete dipole approximation (DDA) simulation method to estimate the plasmon shifts as a function of the separation between the nanoparticles \cite{cpm2}. However, the developed models based on quantum theories are neither lack of details on the emission spectra, especially for PL, nor lack of intrinsic physical pictures. Hence, a clear picture for coupling PL spectra is required to be built up.
%the corresponding experiments are lack of explanations with detailed physics, most of which are either phenomenological explained in the view of quantum transition among the energy levels, which could hardly tell the emission details of PL, or lack of physical formulas.

In this study, we present a practical model to give a deep understanding on PL from coupled metallic nanostructures, e.g., gold nanorods or nanospheres. This model is based on the classic harmonic oscillator model, considering two oscillators that interact with each other. We treat the interaction part, i.e., coupling coefficients, between the two in a non-phenomenological way. That is, the coupling coefficients are obtained from the intrinsic physics rather than just assuming as several parameters. The model show reasonable results to explain PL and white light scattering spectra of coupled metallic nanostructures for different situations. This work would help to understand coupling PL phenomena in a classical way.

~\\ \indent
%\section{\label{sec:Model}Model}
%\textbf{\textit{Model}}.---
Since there are plenty free electrons in the metallic nanostructure, and these electrons oscillate when excited by the external electric field, we treat the nanostructure as a resonator, the oscillators of which are the electrons. Due to the collectively oscillating, we can simplify the multiple electrons as only one electron.
\begin{figure}[tb]
\includegraphics[width=0.48\textwidth]{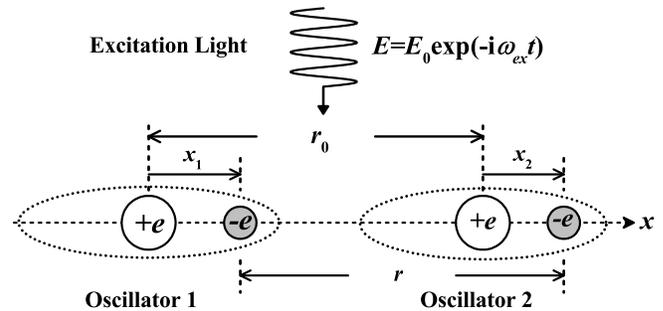}
\caption{\label{fig:Schematic} Schematic of the coupling harmonic oscillator model. The electrons (grey, negative charged) oscillate collectively along $x$-axis near their equilibrium positions. The ions (white, positive charged) is at rest. $r_0$ is the distance between the two ions, while $r$ is the distance between the two electrons. $x_1$ and $x_2$ are the displacements relative to equilibrium positions of
each oscillator. The two oscillators both oscillate along $x$-axis when excited by the excitation light at the circular frequency of $\omega_{ex}$ which is $x$-polarized.
}
\end{figure}
Consider that two metallic nanostructures treated as two oscillators are close to each other and are driven by the electric field of the excitation light. The schematic is shown in Fig. \ref{fig:Schematic}. In order to obtain the emission electric field from them, we need to find out the differential equations of them. Define $x_1(t)$ and $x_2(t)$ as the displacements relative to equilibrium positions of each oscillator, thus $\dot{x}_1(t)$ and $\dot{x}_2(t)$ the velocities, and $\ddot{x}_1(t)$ and $\ddot{x}_2(t)$ the accelerations. The equations should be in this form:
\begin{subequations}
\begin{align}
\ddot{x}_1+2 \beta_{01} \dot{x}_1+\omega_{01}^2 x_1-\frac{F_{21}}{m_e}=C_1 \mathrm{exp}(-\mathrm{i} \omega_{ex} t), \label{eq:basic01a} \\
\ddot{x}_2+2 \beta_{02} \dot{x}_2+\omega_{02}^2 x_2-\frac{F_{12}}{m_e}=C_2 \mathrm{exp}(-\mathrm{i} \omega_{ex} t). \label{eq:basic01b}
\end{align}
\label{eq:basic01}
\end{subequations}
Here, $F_{21}$ and $F_{12}$ are the interaction forces between Oscillator 1 and Oscillator 2, $C_1=E_1/m_e$, $C_2=E_2/m_e$, and $m_e$ is the mass of electron. $E_1$ and $E_2$ are the amplitudes of the excitation electric field at the positions of the two oscillators, and usually $E_1=E_2=E_0$ is a good approximation. $\beta_{01}$ and $\beta_{02}$ represent the damping coefficients, and $\omega_{01}$ and $\omega_{02}$ represent the inherent circular frequencies. The next step is to find out the interaction parts of the equations.

The electric field introduced by a moving charged particle is given by:\cite{Griffiths}
\begin{equation}
\mathbf{E}=\frac{q}{4 \pi \varepsilon_0} \frac{r}{(\mathbf{r} \cdot \mathbf{u})^3} \left [ (c^2-v^2)\mathbf{u} + \mathbf{r} \times (\mathbf{u} \times \mathbf{a}) \right ],
\label{eq:E}
\end{equation}
where $q$ is the charge of the particle, $\varepsilon_0$ is the permittivity of vacuum, and $\mathbf{u} \equiv c \mathbf{r} /r - \mathbf{v}$. Here, $c$ is the velocity of light in vacuum, $\mathbf{v}$ and $\mathbf{a}$ are the velocity and the acceleration of the particle, respectively, and $\mathbf{r}$ is the displacement vector from the particle to field point. In our one-dimension case, when considering the electric field introduced by one oscillator acting on the other oscillator, the second part of Eq. \ref{eq:E} is zero due to the fact that $\mathbf{u}$ and $\mathbf{a}$ are parallel. Besides, we notice that the charged particle that moves is the electron while the positive ion is assumed to be at rest. Hence, the interacted electric field at one oscillator should be contributed to both positive charged ion and negative charged electron of the other oscillator. Therefore, the electric field can be written as:
\begin{subequations}
\begin{align}
E_{21}&= - \frac{+e}{4 \pi \varepsilon_0 } \frac{1}{(r_0-x_1)^2} + \frac{-e}{4 \pi \varepsilon_0} \frac{1}{r^2} (-\frac{c-\dot{x}_2}{c+\dot{x}_2})
\notag
\\ & \cong -\frac{e}{2 \pi \varepsilon_0 r_0^2}(\frac{\dot{x}_2}{c}+\frac{x_2}{r_0}), \label{eq:E2a} \\
E_{12}&=+ \frac{+e}{4 \pi \varepsilon_0 } \frac{1}{(r_0+x_2)^2} + \frac{-e}{4 \pi \varepsilon_0} \frac{1}{r^2} (\frac{c+\dot{x}_2}{c-\dot{x}_2})
\notag
\\& \cong -\frac{e}{2 \pi \varepsilon_0 r_0^2}(\frac{\dot{x}_1}{c}+\frac{x_1}{r_0}). \label{eq:E2b}
\end{align}
\label{eq:E2}
\end{subequations}
Here, we use the conditions $v/c \ll 1$ and $x/r_0 \ll 1$ for approximation. Notice that $E_{21}$ is the electric field in Oscillator 1 introduced by one pair of the electrons and ions in Oscillator 2, and $E_{12}$ is the electric field in Oscillator 2 introduced by one pair of the electrons and ions in Oscillator 1. Hence, the interaction forces should be written as $F_{21}=-N_2 e E_{21}$ and $F_{12}=-N_1 e E_{12}$, where $N_1$ and $N_2$ are the effective numbers of free electrons in Oscillator 1 and Oscillator 2, respectively. We define the coupling coefficients as:
\begin{equation}
\begin{aligned}
\gamma_{21}=\frac{N_2 e^2}{2 \pi \varepsilon_0 m_e r_0^2 c},\ \gamma_{12}=\frac{N_1 e^2}{2 \pi \varepsilon_0 m_e r_0^2 c}, \\
g_{21}^2=\frac{N_2 e^2}{2 \pi \varepsilon_0 m_e r_0^3},\ g_{12}^2=\frac{N_1 e^2}{2 \pi \varepsilon_0 m_e r_0^3},
\end{aligned}
\label{eq:g}
\end{equation}
thus Eq. (\ref{eq:basic01}) can be written as:
\begin{subequations}
\begin{align}
\ddot{x}_1+2 \beta_{01} \dot{x}_1+\omega_{01}^2 x_1-\gamma_{21} \dot{x}_2-g_{21}^2 x_2=C_1 \mathrm{exp}(-\mathrm{i} \omega_{ex} t), \label{eq:basic1a} \\
\ddot{x}_2+2 \beta_{02} \dot{x}_2+\omega_{02}^2 x_2-\gamma_{12} \dot{x}_1-g_{12}^2 x_1=C_2 \mathrm{exp}(-\mathrm{i} \omega_{ex} t). \label{eq:basic1b}
\end{align}
\label{eq:basic1}
\end{subequations}
For simplicity, we assume $N_1=N_2=N$, thus $\gamma_{21}=\gamma_{12}=\gamma$, $g_{21}=g_{12}=g$, and if we define $ \frac{1}{\kappa} = \frac{\gamma^3}{g^4}=\frac{N e^2}{2 \pi \varepsilon_0 m_e c^3} $, it results in a simple form for $\gamma$ and $g$:
\begin{equation}
\gamma=\frac{1}{\kappa} (\frac{c}{r_0})^2,\ g^2=\frac{1}{\kappa} (\frac{c}{r_0})^3.
\label{eq:gammag}
\end{equation}

Firstly, we consider the situation without coupling, i.e., $\gamma=g=0$. In such conditions, Eq. \ref{eq:basic1} is degenerated into the simple form:
\begin{subequations}
\begin{align}
\ddot{x}_1+2 \beta_{01} \dot{x}_1+\omega_{01}^2 x_1=C_1 \mathrm{exp}(-\mathrm{i} \omega_{ex} t), \label{eq:basic2a} \\
\ddot{x}_2+2 \beta_{02} \dot{x}_2+\omega_{02}^2 x_2=C_2 \mathrm{exp}(-\mathrm{i} \omega_{ex} t). \label{eq:basic2b}
\end{align}
\label{eq:basic2}
\end{subequations}
The general solutions are:
\begin{subequations}
\begin{align}
x_1(t)=\mathrm{exp}(-\beta_{01} t \pm \mathrm{i} \omega_{c1} t),\ \mathrm{exp}(-\mathrm{i} \omega_{ex} t), \label{eq:s2a} \\
x_2(t)=\mathrm{exp}(-\beta_{02} t \pm \mathrm{i} \omega_{c2} t),\ \mathrm{exp}(-\mathrm{i} \omega_{ex} t). \label{eq:s2b}
\end{align}
\label{eq:s2}
\end{subequations}
Here, $\omega_{c1}=\sqrt{\omega_{01}^2-\beta_{01}^2}$ and $\omega_{c2}=\sqrt{\omega_{02}^2-\beta_{02}^2}$ represent the resonant circular frequencies, respectively, which are different from the inherent ones ($\omega_{01},\ \omega_{02}$). The coefficients that represent amplitudes are omitted for the moment, which can be obtained with the initial conditions. The details of this kind of individual oscillator has been discussed carefully in our previous work. \cite{PL0}

Secondly, we start to consider the coupling situation without excitation light, i.e., $C_1=C_2=0$. The equations are:
\begin{subequations}
\begin{align}
\ddot{x}_1+2 \beta_{01} \dot{x}_1+\omega_{01}^2 x_1-\gamma \dot{x}_2 -g^2 x_2 =0, \label{eq:basic3a} \\
\ddot{x}_2+2 \beta_{02} \dot{x}_2+\omega_{02}^2 x_2-\gamma \dot{x}_1 -g^2 x_1 =0. \label{eq:basic3b}
\end{align}
\label{eq:basic3}
\end{subequations}
To solve Eq. (\ref{eq:basic3}), we can assume that $x_1(t)=A \mathrm{exp}(\alpha t)$ and $x_2(t)=B \mathrm{exp}(\alpha t)$, and substitute them back into Eq. (\ref{eq:basic3}), thus obtaining:
\begin{subequations}
\begin{align}
A(\alpha^2 + 2 \beta_{01} \alpha + w_{01}^2) - B (\gamma \alpha + g^2)=0, \label{eq:alpha1a} \\
B(\alpha^2 + 2 \beta_{02} \alpha + w_{02}^2) - A (\gamma \alpha + g^2)=0. \label{eq:alpha1b}
\end{align}
\label{eq:alpha1}
\end{subequations}
Obviously, to obtain non-zero solutions, $\alpha$ should satisfy:
\begin{equation}
(\alpha^2 + 2 \beta_{01} \alpha + w_{01}^2)(\alpha^2 + 2 \beta_{02} \alpha + w_{02}^2)= (\gamma \alpha + g^2)^2.
\label{eq:alpha2}
\end{equation}
Notice that Eq. (\ref{eq:alpha2}) has analytic solutions for $\alpha$, marked as $\alpha_1$, $\alpha_2$, $\alpha_3$ and $\alpha_4$. However, the expressions are so complex that we would not write in the text. Instead, to illustrate the physical significance of $\alpha$, we rewrite it in this form:
\begin{equation}
\begin{aligned}
\alpha_1=-\beta_1 + \mathrm{i} \omega_1,\ \alpha_2=-\beta_1 - \mathrm{i} \omega_1,\\
\alpha_3=-\beta_2 + \mathrm{i} \omega_2,\
\alpha_4=-\beta_2 - \mathrm{i} \omega_2.
\end{aligned}
\label{eq:alpha3}
\end{equation}
Here, $\omega_1$ and $\omega_2$ are the new generated resonant circular frequencies when the two oscillators couple. We can call them Mode 1 and Mode 2, respectively.
In a more special case, i.e., $\beta_{01}=\beta_{02}=\beta_0$, $\omega_{01}=\omega_{02}=\omega_0$, the solutions of Eq. (\ref{eq:alpha2}) are expressed easily:
\begin{equation}
\begin{aligned}
\omega_1=\sqrt{w_0^2+g^2-(\beta_0 + \gamma/2)^2},\ \beta_1=\beta_0 + \gamma/2,\\
\omega_2=\sqrt{w_0^2-g^2-(\beta_0 - \gamma/2)^2},\ \beta_2=\beta_0 - \gamma/2.
\end{aligned}
\label{eq:wb}
\end{equation}

Thirdly, notice that the particular solutions for Eq. (\ref{eq:basic1}) are $x_1(t)=\mathrm{exp}(-\mathrm{i} \omega_{ex} t)$ and $x_2(t)=\mathrm{exp}(-\mathrm{i} \omega_{ex} t)$ (amplitudes are omitted). Therefore, combining these particular solutions and the general ones [Eq. (\ref{eq:s2})], we obtain the total solutions of Eq. (\ref{eq:basic1}) in a symmetric form:
\begin{subequations}
\begin{align}
x_1(t)&=A_1 \mathrm{exp}(\Omega_1 t) + A_2 \mathrm{exp}(\Omega_2 t) + A_3 \mathrm{exp}(\Omega_3 t), \label{eq:s1a}\\
x_2(t)&=B_1 \mathrm{exp}(\Omega_1 t) + B_2 \mathrm{exp}(\Omega_2 t) + B_3 \mathrm{exp}(\Omega_3 t), \label{eq:s1b}
\end{align}
\label{eq:s1}
\end{subequations}
where $\Omega_1=-\beta_1 - \mathrm{i} \omega_1$, $\Omega_2=-\beta_2 - \mathrm{i} \omega_2$, and $\Omega_3= - \mathrm{i} \omega_{ex}$.
We emphasize here that Eq. (\ref{eq:wb}) is just a special case for $\omega_1$ and $\omega_2$, and the general case for them should satisfy Eq. (\ref{eq:alpha3}).
The initial conditions are $x_1(0)=x_2(0)=0,\ \dot{x}_1(0)=\dot{x}_2(0)=0,\ \ddot{x}_1(0)=\ddot{x}_2(0)=C_0$, where we assume that $C_1=C_2=C_0$ due to the subwavelength scale of the system. Hence, these coefficients are obtained as:
\begin{subequations}
\begin{align}
A_1=B_1=\frac{C_0}{(\Omega_1-\Omega_2)(\Omega_1-\Omega_3)},
\label{eq:ampa}\\
A_2=B_2=\frac{C_0}{(\Omega_2-\Omega_3)(\Omega_2-\Omega_1)},
\label{eq:ampb}\\
A_3=B_3=\frac{C_0}{(\Omega_3-\Omega_1)(\Omega_3-\Omega_2)}.
\label{eq:ampc}
\end{align}
\label{eq:amp}
\end{subequations}
This results in the fact that $x_1(t)=x_2(t)=x(t)$.

At last, we deal with the far field radiation. For simplicity, we consider the electric field at the position $\mathbf{d}$, where $\mathbf{d}$ is perpendicular to $x$-axis, and $d=|\mathbf{d}|$ is the distance between field point and the center of the two oscillators. The assumption of $d \gg r_0$ is reasonable for far field radiation. Hence, the first part of Eq. (\ref{eq:E}) is ignored compared with the second part, thus giving the electric field introduced by Oscillator 1 and Oscillator 2 as:
\begin{equation}
E_{far}(t) \cong \frac{N e}{4 \pi \varepsilon_0 c^2 d} (\ddot{x}_1 (t)+\ddot{x}_2 (t))=D \ddot{x} (t),
\label{eq:Efar}
\end{equation}
where $D=\frac{N e}{2 \pi \varepsilon_0 c^2 d}$, and $E_{far}$ is $x$-polarized. The emission intensity in the frequency domain, i.e., emission spectrum, can be evaluated by \cite{PL0}:
\begin{equation}
I(\omega)=Re \left < \int_0^{\infty} E_{far}^*(t) E_{far}(t+\tau)\  \mathrm{exp}(\mathrm{i} \omega \tau)\ \mathrm{d} \tau   \right >,
\label{eq:I1}
\end{equation}
where $Re \left < Q \right >$ is the real part of $ \left < Q \right >$, and $ \left < Q \right > = \frac{1}{t_0}\int_0^{t_0} Q\mathrm{d} t$ is the time average of quantity $Q$.  The calculated result is:
\begin{equation}
\begin{aligned}
I(\omega)&=|A_1'|^2 \frac{1-\mathrm{exp}(-2 \beta_1 t_0)}{2 \beta_1 t_0} \frac{\beta_1}{(\omega - \omega_1)^2+\beta_1^2}
\\ & + |A_2'|^2 \frac{1-\mathrm{exp}(-2 \beta_2 t_0)}{2 \beta_2 t_0} \frac{\beta_2}{(\omega - \omega_2)^2+\beta_2^2}
\\ & + |A_3'|^2 \sqrt{2 \pi} \delta(\omega-\omega_{ex}),
\end{aligned}
\label{eq:I2}
\end{equation}
where $A_j'=A_j \Omega_j^2 D$ for $j=1,2,3$. Here, we ignore the cross terms in the calculation because the time average is zero when $\omega_1 \neq \omega_2$.

As our previous work explains \cite{PL0}, the emission spectrum is separated into two parts, one is the inelastic part ($I_{inela}$) which corresponds to PL spectrum, and the other is the elastic part ($I_{ela}$) which corresponds to white light scattering spectrum. Rewrite Eq. (\ref{eq:I2}) as:
\begin{subequations}
\begin{align}
I_{inela}(\omega)&=|A_1'|^2 \frac{1-\mathrm{exp}(-2 \beta_1 t_0)}{2 \beta_1 t_0} \frac{\beta_1}{(\omega - \omega_1)^2+\beta_1^2}
\notag
\\&  + |A_2'|^2 \frac{1-\mathrm{exp}(-2 \beta_2 t_0)}{2 \beta_2 t_0} \frac{\beta_2}{(\omega - \omega_2)^2+\beta_2^2}, \label{eq:separatea}
\\ I_{ela}(\omega)&=|A_3'|^2 \sqrt{2 \pi} \delta(\omega-\omega_{ex}). \label{eq:separateb}
\end{align}
\label{eq:separate}
\end{subequations}
Therefore, the PL spectrum is given by Eq. (\ref{eq:separatea}), i.e.,
\begin{equation}
I_{PL}(\omega)=I_{inela}(\omega),
\label{eq:PL}
\end{equation}
while the white light scattering spectrum is given from Eq. (\ref{eq:separateb}) as long as $\omega_{ex}$ is substitute by $\omega$:
\begin{equation}
I_{sca}(\omega)=I_{ela}(\omega_{ex} \to \omega)=\sqrt{2 \pi} |A_3'(\omega_{ex} \to \omega)|^2 .
\label{eq:sca}
\end{equation}

To show the coupling modes for PL more clearly and to understand PL phenomenon more easily, we do not consider the electron distributions here as before \cite{PL0}, which contributes mostly to the anti-Stokes part of PL spectra, though this model would be more accuracy for PL when assisted with the electron distributions.

~\\ \indent
%\section{\label{sec:Results}Results and Discussions}
After obtaining these formulas, we would analyze in details to understand them more deeply.

\begin{figure}[tb]
\includegraphics[width=0.48\textwidth]{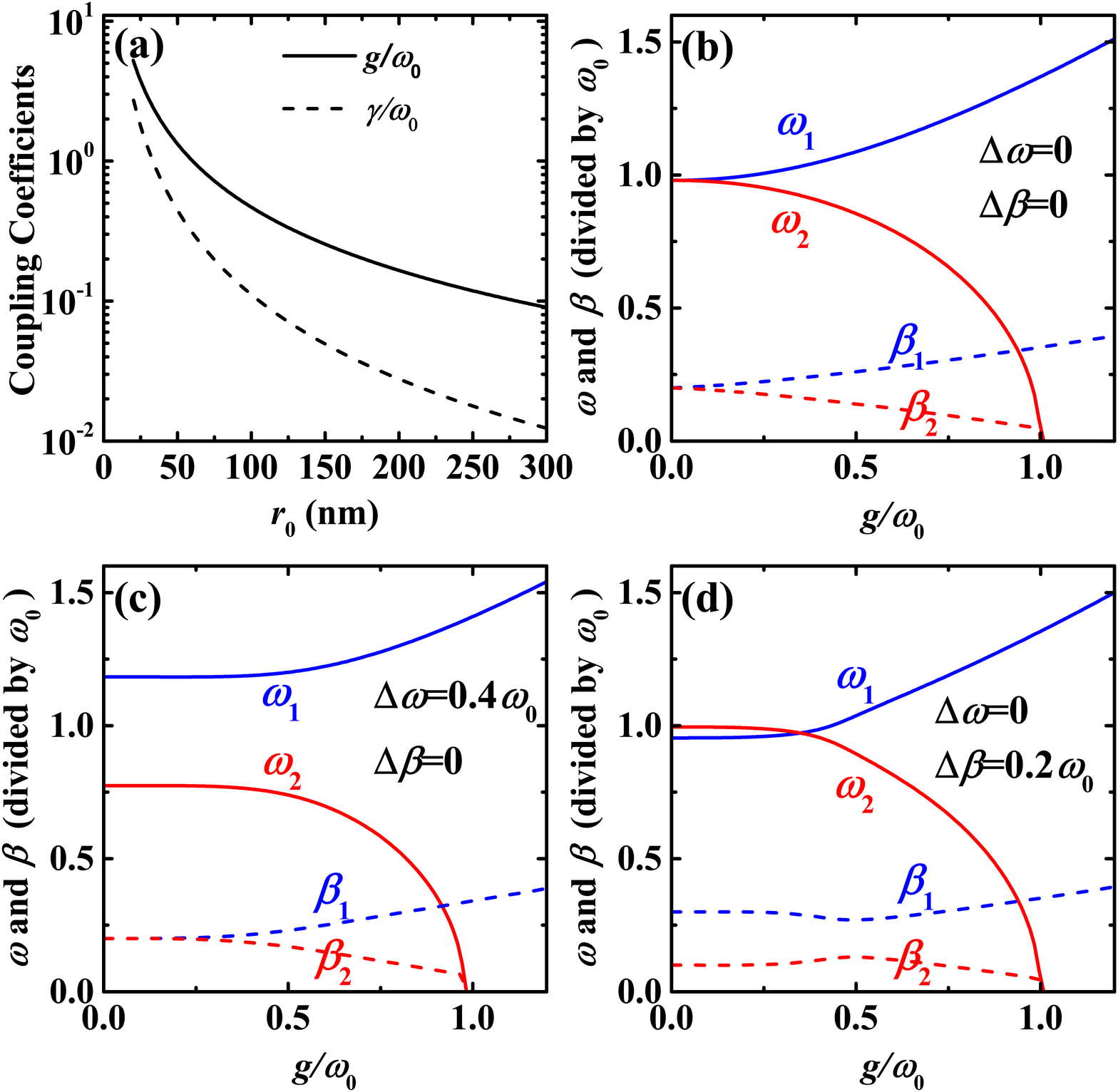}
\caption{\label{fig:omegabeta} (a) Coupling coefficients $g$ (solid curve) and $\gamma$ (dashed curve) as a function of $r_0$. (b), (c), (d) The new generated resonant circular frequencies ($\omega_1,\ \omega_2$) and damping coefficients ($\beta_1,\ \beta_2$) as a function of $g$. Here, $\omega_{01}=\omega_0+\Delta \omega /2,\ \omega_{02}=\omega_0-\Delta \omega /2$, $\beta_{01}=\beta_0+\Delta \beta /2,\ \beta_{02}=\beta_0-\Delta \beta /2$, and $\beta_0=0.2 \omega_0$. The number of effective free electrons is estimated as $N = 10^6$.
}
\end{figure}

Start from the coupling coefficients, $g$ and $\gamma$. Fig. \ref{fig:omegabeta}a shows $g$ and $\gamma$ varying with the distance $r_0$, calculated from Eq. (\ref{eq:gammag}). It implicates that the coupling coefficients decrease with the increase of $r_0$, and $\gamma$ is smaller than $g$. When $r_0$ is small enough, the coupling coefficients get large. Since these two coefficients are both related to $r_0$, we take one of them, i.e., $g$, as the coupling strength in the rest of this work.
Fig. \ref{fig:omegabeta}b-d show the new generated resonant circular frequencies ($\omega_1,\ \omega_2$) and damping coefficients ($\beta_1,\ \beta_2$) in different cases of the coupled oscillators as a function of the coupling strength $g$, calculated from Eq. (\ref{eq:alpha2}) and (\ref{eq:alpha3}). The simplest one (Fig. \ref{fig:omegabeta}b) is that the two oscillators are the same. The two new modes split when coupling, and the splitting increases with the increase of $g$. Here, we generally call the increasing $\omega$ ``blue branch'', and the decreasing $\omega$ ``red branch''.  On the other hand, the two damping coefficients also splits, and one increases (corresponding blue branch), the other decreases (red branch). Notice that there is a cut-off coupling strength for the red branch at around $g_{cut} \approx \omega_0$. In Fig. \ref{fig:omegabeta}c, the situation is almost the same, i.e., the difference of $\omega$ and $\beta$ between the two branches increase with the increase of $g$, and there also exists $g_{cut}$. The difference between Fig. \ref{fig:omegabeta}b and \ref{fig:omegabeta}c is that, to obtain the same level of splitting, the former needs a smaller $g$ than the latter does. That is, the former gets a better coupling efficient than the latter does. In Fig. \ref{fig:omegabeta}d, due to the fact that $\omega_{c1}=\sqrt{\omega_{01}^2-\beta_{01}^2}=0.954\omega_0$ and $\omega_{c2}=\sqrt{\omega_{02}^2-\beta_{02}^2}=0.995\omega_0$, it results in $\omega_{c1} < \omega_{c2}$ with a small difference. The difference of $\omega$ between the two branches ($\omega_1-\omega_2$) increases from negative value to zero and then increase to positive value as the increase of $g$. On the other hand, the difference of $\beta$ between the two branches ($\beta_1-\beta_2$) decreases and then increases as the increase of $g$. Also, $g_{cut}$ exists in this case. The coupling efficient in Fig. \ref{fig:omegabeta} follows the relation: $\mathrm{(b)} > \mathrm{(d)} > \mathrm{(c)}$.

\begin{figure}[tb]
\includegraphics[width=0.5\textwidth]{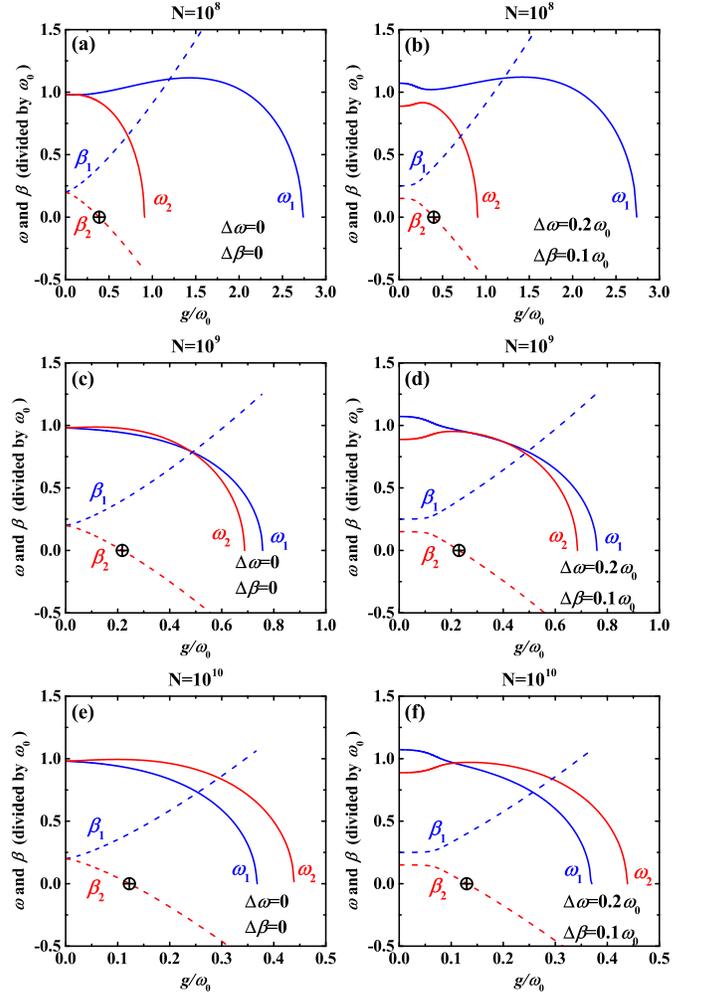}
\caption{\label{fig:omegabeta2} The new generated resonant circular frequencies ($\omega_1,\ \omega_2$) and damping coefficients ($\beta_1,\ \beta_2$) as a function of $g$, varying with effective free electrons number $N$. (a), (c) and (e) represent the case of two same oscillators, where $\omega_{01}=\omega_{02}=\omega_0$ and $\beta_{01}=\beta_{02}=0.2\omega_0$. (b), (d) and (f) represent the case of two different oscillators, with $\Delta \omega=0.2\omega_0$ and $\Delta \beta=0.1\omega_0$. (a) and (b), (c) and (d), (e) and (f)  represent $N=10^8$, $N=10^9$, $N=10^{10}$, respectively. The black cross circles represent the point at which $\beta_2=0$.
}
\end{figure}
Furthermore, the effective free electrons number affects the splitting as shown in Fig. \ref{fig:omegabeta2}, giving three values of $N$ as examples.
Firstly, we find out the behaviors of $\omega_1$ and $\omega_2$ as $g$ increases for each figure. In Fig. \ref{fig:omegabeta2}a, $\omega_1$ increases and then decreases, while $\omega_2$ decreases, indicating that $\omega_1$ has a maximal value. In Fig. \ref{fig:omegabeta2}b, $\omega_1$ decreases and then increases and finally decreases, while $\omega_2$ increases and then decreases, indicating that both $\omega_1$ and $\omega_2$ have maximal values. In Fig. \ref{fig:omegabeta2}c, $\omega_1$ decreases, while $\omega_2$ increases slightly and then decreases. In Fig. \ref{fig:omegabeta2}d, $\omega_1$ decreases, while $\omega_2$ increases and then decreases, the curves of which almost coincide with each other at the range around $g=0.3\omega_0$ to $g=0.5\omega_0$. In Fig. \ref{fig:omegabeta2}e, the behaviors are similar to the ones in Fig. \ref{fig:omegabeta2}c. In Fig. \ref{fig:omegabeta2}f, the behaviors are similar to the ones in Fig. \ref{fig:omegabeta2}d, but the two curves cross rather than coincide.
Secondly, we find out the similar behaviors for these parameters in a general view. In all cases, there are cut-off coupling strengths for both modes, writing as $g_{cut1}$ and $g_{cut2}$, at which $\omega_1=0$ and $\omega_2=0$, respectively. The differences are, for smaller $N$ ($10^8$ or $10^9$), $g_{cut1}>g_{cut2}$, while for larger $N$ ($10^{10}$), $g_{cut1}<g_{cut2}$. As $g$ increases, the splitting of damping coefficients $\beta_1$ and $\beta_2$ gets larger. Furthermore, another interesting result is that there is a point $g_0$ (shown with black cross circle) at which $\beta_2 (g_0)=0$ for each case, and $g_0<g_{cut2}$. This is different from the one in Fig. \ref{fig:omegabeta} where $g_0>g_{cut2}$. When $g<g_0$, Mode 2 behaves normally. However, when $g>g_0$, $\beta_2<0$ indicates that this is an exponentially increasing mode, which should be removed from the total solutions [Eq. (\ref{eq:s1})], resulting in the absence of Mode 2. The most special case is when $g=g_0$ (or $g \to g_0^+$), which corresponds to a lossless (or low loss) mode. In frequency domain this mode would results in a narrow spectrum.
However, the effective free electrons number that satisfy this condition is so large that it is almost impossible for a metallic nanostructure. Therefore, in the rest of this work, we only consider the number at the order of magnitudes of $N=10^6$.

In Eq. (\ref{eq:s1}), $A_1$, $A_2$ and $A_3$ represent the amplitudes of the three corresponding modes of $x(t)$. When considering the far field, one should use the amplitudes of $\ddot{x}(t)$, i.e., $A_1'$, $A_2'$ and $A_3'$. Obviously, the frequency of the excitation light plays a significant role in the amplitudes.
\begin{figure}[tb]
\includegraphics[width=0.48\textwidth]{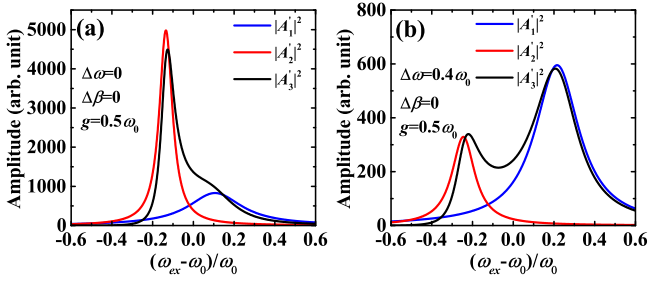}
\caption{\label{fig:Amplitude}  Amplitudes of modes $\omega_1$ ($|A_1'|^2$), $\omega_2$ ($|A_2'|^2$) and $\omega_{ex}$ ($|A_3'|^2$) as a function of $\omega_{ex}$.
(a) $\Delta \omega =0$, $\Delta \beta =0$. (b) $\Delta \omega =0.4 \omega_0$, $\Delta \beta =0$.
The definitions of $\Delta \omega$ and $\Delta \beta$ are the same as Fig. \ref{fig:omegabeta} except for $\beta_0=0.1\omega_0$.
Here, the coupling strength is $g=0.5\omega_0$, and $C_0=1$ and $D=1$ are used for normalization.
}
\end{figure}
Fig. \ref{fig:Amplitude} shows these amplitudes as a function of $\omega_{ex}$. In the first case (Fig. \ref{fig:Amplitude}a), i.e., two same oscillators, the coupled resonant circular frequencies (relative to $\omega_0$) are calculated as $(\omega_1-\omega_0)/\omega_0=0.11$ and $(\omega_2-\omega_0)/\omega_0=-0.13$. We find that to obtain the maximum intensities of Mode 1 and Mode 2 of the emission field, the circular frequency of the excitation light $\omega_{ex}$ should be close the corresponding circular resonant frequencies. For Mode 3, there are two peaks when varying $\omega_{ex}$, which correspond to around $\omega_1$ and $\omega_2$, respectively. In the second case (Fig. \ref{fig:Amplitude}b), i.e., two different oscillators ($\omega_1 > \omega_2$), the coupled resonant circular frequencies (relative to $\omega_0$) are calculated as 0.22 and -0.24, respectively, which, however, corresponds to a weak coupling due to the frequency splitting is small. This result has been identified in Fig. \ref{fig:omegabeta}. Also, the intensities of Mode 1 and Mode 2 for far field reach their maximums when $\omega_{ex}$ is close to the resonant circular frequencies for each of them, and the two corresponding peaks appear for Mode 3.

\begin{figure}[tb]
\includegraphics[width=0.48\textwidth]{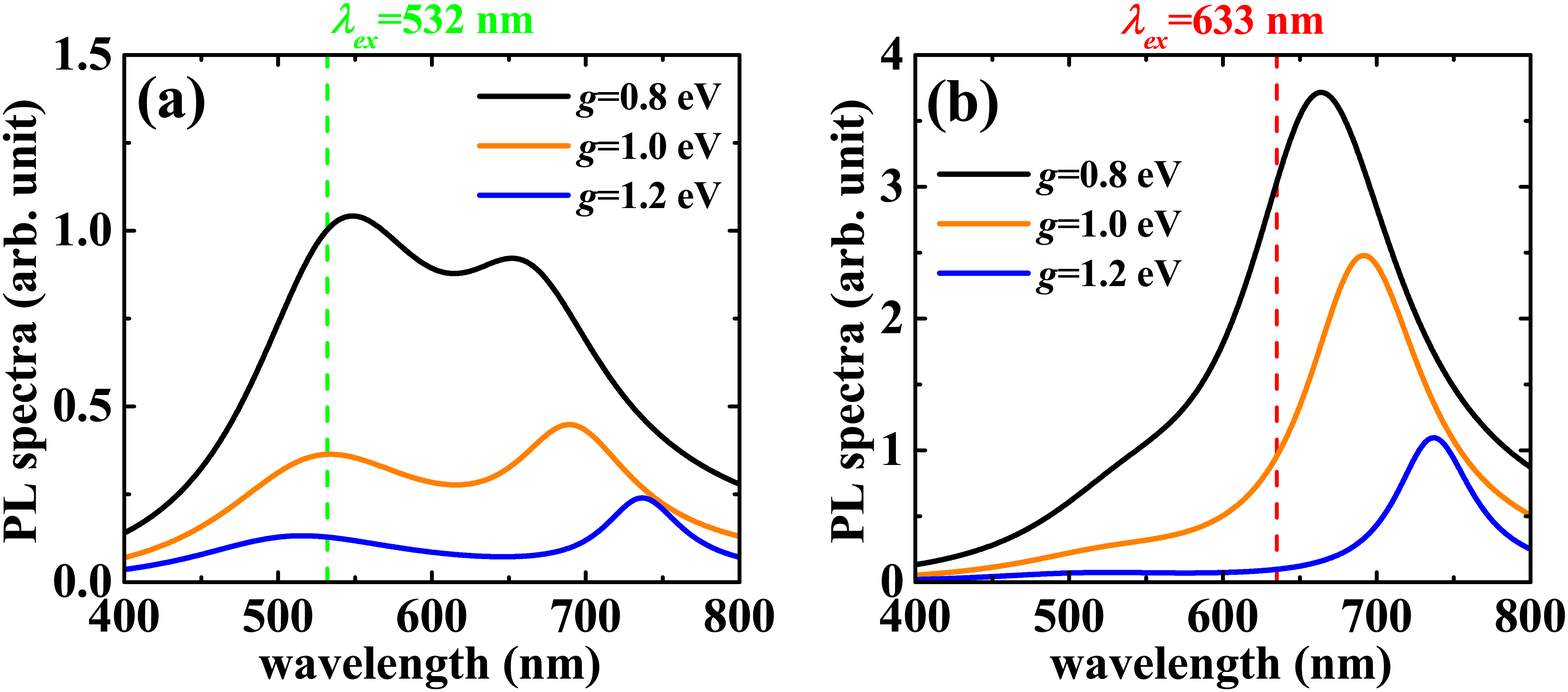}
\caption{\label{fig:PL}  PL spectra of the two coupled oscillators at $g=0.8$ eV (black), 1.0 eV (orange) and 1.2 eV (blue), respectively, calculated from Eq. (\ref{eq:PL}). The excitation light is at the wavelength of $\lambda_{ex}=532$ nm (a) and $\lambda_{ex}=633$ nm (b). Here, $\lambda_{c1}=550$ nm and $\lambda_{c2}=650$ nm represent the resonant wavelengths for each oscillator (before coupling), respectively; $\beta_{01}=\beta_{02}=0.247$ eV. Vertical dashed lines stand for the position of 532 nm (green) and 633 nm (red), respectively.
}
\end{figure}

For practical purpose, we consider two metallic nanostructures, e.g., gold nanorods or nanospheres, as the two oscillators, each of which has an individual resonant mode. Fig. \ref{fig:PL} shows the coupling PL spectra for different coupling strengths at two different excitation wavelengths, calculated from Eq. (\ref{eq:PL}). With the increase of $g$, the splitting of the two modes of PL increases, and  the total emission intensities decrease. The decrease of the intensities origin from Eq. (\ref{eq:ampa}) and (\ref{eq:ampb}). Take Eq. (\ref{eq:ampa}) as an example to explain. The amplitudes depend not only on $|\omega_1-\omega_{ex}|$ (this has been discussed in Fig. \ref{fig:Amplitude}), but also on $|\omega_1-\omega_2|$. When $g$ increases, $|\omega_1-\omega_2|$ increases, resulting in the decrease of the amplitude of Mode 1. So does Mode 2. Therefore, the PL intensities decrease as $g$ increases.
Besides, when excited by 532 nm laser, Mode 1 is close to it, resulting in a larger intensity than the one of Mode 2. While excited by 633 nm laser, Mode 2 is close to it, resulting in a larger intensity than the one of Mode 1. This is consistent with the results in Fig. \ref{fig:Amplitude}. %and the experiments \cite{}.
Here, unit ``eV'' and unit ``Hz'' for $g$ satisfy the following relationship:
\begin{equation}
g[\mathrm{eV}]=\frac{\hbar}{e}g[\mathrm{Hz}],
\label{eq:eV}
\end{equation}
where $\hbar$ is the reduced Planck constant. So does the damping coefficient $\beta$.

\begin{figure}[tb]
\includegraphics[width=0.48\textwidth]{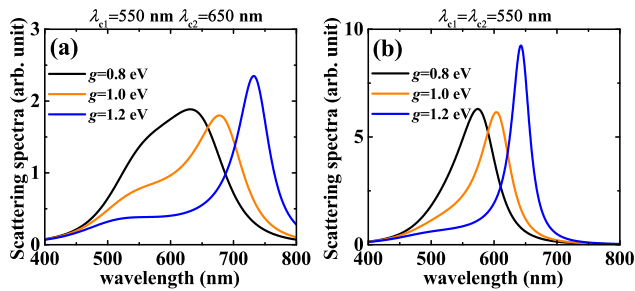}
\caption{\label{fig:Sca} White light scattering spectra of the two coupled oscillators at $g=0.8$ eV (black), 1.0 eV (orange) and 1.2 eV (blue), respectively, calculated from Eq. (\ref{eq:sca}). (a) The resonant wavelengths are different, $\lambda_{c1}=550$ nm and $\lambda_{c2}=650$ nm, respectively. (b) The resonant wavelengths are the same, $\lambda_{c1}=\lambda_{c2}=550$ nm. Here, the damping coefficients are the same for all the oscillators, $\beta_0=0.247$ eV .
}
\end{figure}
Fig. \ref{fig:Sca} shows the coupling white light scattering spectra for different coupling strengths in different cases, calculated from Eq. (\ref{eq:sca}). In Fig. \ref{fig:Sca}a, i.e., two oscillators with different resonant wavelengths, with the increase of $g$, the splitting of the two modes increases, which behaves the same as PL does. However, the scattering intensities stay in the same level which is different from PL spectra. In Fig. \ref{fig:Sca}b, i.e., two same oscillators, with the increase of $g$, Mode 2 red-shifts, while Mode 1 is hardly to be obtained. Also, the intensities  stay in the same level. This behavior agrees well with the experiments \cite{cp1,cp2,cp3}.
%in the latter of which two individual 80 nm gold nanodisks are coupled with varied gap size.

~\\ \indent
%\section*{\label{sec:Conclusion}Conclusions}
In summary, we develop a coupling classic harmonic oscillator model to explain the coupling PL spectra as well as the white light scattering spectra from two coupled metallic nanostructures. Each nanostructure is treated as a classic charged oscillator with its own single mode. The coupling coefficients are obtained from the electric interactions between the charges, and are proportional to the velocity and the acceleration of the oscillator, respectively. The behaviors of the two new generated modes due to the coupling are different under different conditions. In general, they split and the splitting gets large as the coupling strength $g$ increases at the beginning. Meanwhile, tuning effective free electron number $N$, when $g$ gets large enough, there exist cut-off frequencies for both modes, and a maximum frequency for one of the modes. Besides, PL spectra and white light scattering spectra are calculated from the model, and their behaviors varying with the coupling strength agree well with the experimental ones of other researchers' work. It is worth noting that this coupling model could be expanded to other wavebands dealing with two coupled single-mode resonators.

~\\ \indent% add an empty line
%\section*{\label{sec:Acknowldegment}Acknowledgment}
This work was supported by the Fundamental Research Funds for the Central Universities (Grant No. FRF-TP-20-075A1).

%\section*{Disclosures}
The authors declare no conflicts of interest.

%\section*{Data availability}
The data that support the findings of this study are available from the corresponding author upon reasonable request.

\section*{\label{sec:Ref}References}
%\bibliography{viscosity_Cheng}
%\balance
\bibliography{CM_Cheng}

\begin{thebibliography}{25}%
\makeatletter
\providecommand \@ifxundefined [1]{%
 \@ifx{#1\undefined}
}%
\providecommand \@ifnum [1]{%
 \ifnum #1\expandafter \@firstoftwo
 \else \expandafter \@secondoftwo
 \fi
}%
\providecommand \@ifx [1]{%
 \ifx #1\expandafter \@firstoftwo
 \else \expandafter \@secondoftwo
 \fi
}%
\providecommand \natexlab [1]{#1}%
\providecommand \enquote  [1]{``#1''}%
\providecommand \bibnamefont  [1]{#1}%
\providecommand \bibfnamefont [1]{#1}%
\providecommand \citenamefont [1]{#1}%
\providecommand \href@noop [0]{\@secondoftwo}%
\providecommand \href [0]{\begingroup \@sanitize@url \@href}%
\providecommand \@href[1]{\@@startlink{#1}\@@href}%
\providecommand \@@href[1]{\endgroup#1\@@endlink}%
\providecommand \@sanitize@url [0]{\catcode `\\12\catcode `\$12\catcode
  `\&12\catcode `\#12\catcode `\^12\catcode `\_12\catcode `\%12\relax}%
\providecommand \@@startlink[1]{}%
\providecommand \@@endlink[0]{}%
\providecommand \url  [0]{\begingroup\@sanitize@url \@url }%
\providecommand \@url [1]{\endgroup\@href {#1}{\urlprefix }}%
\providecommand \urlprefix  [0]{URL }%
\providecommand \Eprint [0]{\href }%
\providecommand \doibase [0]{http://dx.doi.org/}%
\providecommand \selectlanguage [0]{\@gobble}%
\providecommand \bibinfo  [0]{\@secondoftwo}%
\providecommand \bibfield  [0]{\@secondoftwo}%
\providecommand \translation [1]{[#1]}%
\providecommand \BibitemOpen [0]{}%
\providecommand \bibitemStop [0]{}%
\providecommand \bibitemNoStop [0]{.\EOS\space}%
\providecommand \EOS [0]{\spacefactor3000\relax}%
\providecommand \BibitemShut  [1]{\csname bibitem#1\endcsname}%
\let\auto@bib@innerbib\@empty
%</preamble>
\bibitem [{\citenamefont {Mooradian}(1969)}]{PL1}%
  \BibitemOpen
  \bibfield  {author} {\bibinfo {author} {\bibfnamefont {A.}~\bibnamefont
  {Mooradian}},\ }\bibfield  {title} {\enquote {\bibinfo {title}
  {Photoluminescence of metals},}\ }\href {\doibase 10.1103/PhysRevLett.22.185}
  {\bibfield  {journal} {\bibinfo  {journal} {Physical Review Letters}\
  }\textbf {\bibinfo {volume} {22}},\ \bibinfo {pages} {185--187} (\bibinfo
  {year} {1969})}\BibitemShut {NoStop}%
\bibitem [{\citenamefont {Boyd}, \citenamefont {Yu},\ and\ \citenamefont
  {Shen}(1986)}]{PL2}%
  \BibitemOpen
  \bibfield  {author} {\bibinfo {author} {\bibfnamefont {G.~T.}\ \bibnamefont
  {Boyd}}, \bibinfo {author} {\bibfnamefont {Z.~H.}\ \bibnamefont {Yu}}, \ and\
  \bibinfo {author} {\bibfnamefont {Y.~R.}\ \bibnamefont {Shen}},\ }\bibfield
  {title} {\enquote {\bibinfo {title} {Photoinduced luminescence from the noble
  metals and its enhancement on roughened surfaces},}\ }\href {\doibase
  10.1103/PhysRevB.33.7923} {\bibfield  {journal} {\bibinfo  {journal}
  {Physical Review B}\ }\textbf {\bibinfo {volume} {33}},\ \bibinfo {pages}
  {7923--7936} (\bibinfo {year} {1986})}\BibitemShut {NoStop}%
\bibitem [{\citenamefont {Mohamed}\ \emph {et~al.}(2000)\citenamefont
  {Mohamed}, \citenamefont {Volkov}, \citenamefont {Link},\ and\ \citenamefont
  {El-Sayed}}]{PL3}%
  \BibitemOpen
  \bibfield  {author} {\bibinfo {author} {\bibfnamefont {M.~B.}\ \bibnamefont
  {Mohamed}}, \bibinfo {author} {\bibfnamefont {V.}~\bibnamefont {Volkov}},
  \bibinfo {author} {\bibfnamefont {S.}~\bibnamefont {Link}}, \ and\ \bibinfo
  {author} {\bibfnamefont {M.~A.}\ \bibnamefont {El-Sayed}},\ }\bibfield
  {title} {\enquote {\bibinfo {title} {The `lightning' gold nanorods:
  fluorescence enhancement of over a million compared to the gold metal},}\
  }\href {\doibase https://doi.org/10.1016/S0009-2614(99)01414-1} {\bibfield
  {journal} {\bibinfo  {journal} {Chemical Physics Letters}\ }\textbf {\bibinfo
  {volume} {317}},\ \bibinfo {pages} {517--523} (\bibinfo {year}
  {2000})}\BibitemShut {NoStop}%
\bibitem [{\citenamefont {Cai}\ \emph {et~al.}(2018)\citenamefont {Cai},
  \citenamefont {Liu}, \citenamefont {Tauzin}, \citenamefont {Huang},
  \citenamefont {Sung}, \citenamefont {Zhang}, \citenamefont {Joplin},
  \citenamefont {Chang}, \citenamefont {Nordlander},\ and\ \citenamefont
  {Link}}]{PL4}%
  \BibitemOpen
  \bibfield  {author} {\bibinfo {author} {\bibfnamefont {Y.-Y.}\ \bibnamefont
  {Cai}}, \bibinfo {author} {\bibfnamefont {J.~G.}\ \bibnamefont {Liu}},
  \bibinfo {author} {\bibfnamefont {L.~J.}\ \bibnamefont {Tauzin}}, \bibinfo
  {author} {\bibfnamefont {D.}~\bibnamefont {Huang}}, \bibinfo {author}
  {\bibfnamefont {E.}~\bibnamefont {Sung}}, \bibinfo {author} {\bibfnamefont
  {H.}~\bibnamefont {Zhang}}, \bibinfo {author} {\bibfnamefont
  {A.}~\bibnamefont {Joplin}}, \bibinfo {author} {\bibfnamefont {W.-S.}\
  \bibnamefont {Chang}}, \bibinfo {author} {\bibfnamefont {P.}~\bibnamefont
  {Nordlander}}, \ and\ \bibinfo {author} {\bibfnamefont {S.}~\bibnamefont
  {Link}},\ }\bibfield  {title} {\enquote {\bibinfo {title} {Photoluminescence
  of gold nanorods: Purcell effect enhanced emission from hot carriers},}\
  }\href {\doibase 10.1021/acsnano.7b07402} {\bibfield  {journal} {\bibinfo
  {journal} {ACS Nano}\ }\textbf {\bibinfo {volume} {12}},\ \bibinfo {pages}
  {976--985} (\bibinfo {year} {2018})}\BibitemShut {NoStop}%
\bibitem [{\citenamefont {Balykin}\ and\ \citenamefont
  {Melentiev}(2018)}]{PL5}%
  \BibitemOpen
  \bibfield  {author} {\bibinfo {author} {\bibfnamefont {V.~I.}\ \bibnamefont
  {Balykin}}\ and\ \bibinfo {author} {\bibfnamefont {P.~N.}\ \bibnamefont
  {Melentiev}},\ }\bibfield  {title} {\enquote {\bibinfo {title} {Optics and
  spectroscopy of a single plasmonic nanostructure},}\ }\href {\doibase
  10.3367/ufne.2017.06.038163} {\bibfield  {journal} {\bibinfo  {journal}
  {Physics-Uspekhi}\ }\textbf {\bibinfo {volume} {61}},\ \bibinfo {pages}
  {133--156} (\bibinfo {year} {2018})}\BibitemShut {NoStop}%
\bibitem [{\citenamefont {Cai}\ \emph {et~al.}(2019)\citenamefont {Cai},
  \citenamefont {Collins}, \citenamefont {Gallagher}, \citenamefont
  {Bhattacharjee}, \citenamefont {Zhang}, \citenamefont {Chow}, \citenamefont
  {Ahmadivand}, \citenamefont {Ostovar}, \citenamefont {Al-Zubeidi},
  \citenamefont {Wang}, \citenamefont {Nordlander}, \citenamefont {Landes},\
  and\ \citenamefont {Link}}]{PL6}%
  \BibitemOpen
  \bibfield  {author} {\bibinfo {author} {\bibfnamefont {Y.-Y.}\ \bibnamefont
  {Cai}}, \bibinfo {author} {\bibfnamefont {S.~S.~E.}\ \bibnamefont {Collins}},
  \bibinfo {author} {\bibfnamefont {M.~J.}\ \bibnamefont {Gallagher}}, \bibinfo
  {author} {\bibfnamefont {U.}~\bibnamefont {Bhattacharjee}}, \bibinfo {author}
  {\bibfnamefont {R.}~\bibnamefont {Zhang}}, \bibinfo {author} {\bibfnamefont
  {T.~H.}\ \bibnamefont {Chow}}, \bibinfo {author} {\bibfnamefont
  {A.}~\bibnamefont {Ahmadivand}}, \bibinfo {author} {\bibfnamefont
  {B.}~\bibnamefont {Ostovar}}, \bibinfo {author} {\bibfnamefont
  {A.}~\bibnamefont {Al-Zubeidi}}, \bibinfo {author} {\bibfnamefont
  {J.}~\bibnamefont {Wang}}, \bibinfo {author} {\bibfnamefont {P.}~\bibnamefont
  {Nordlander}}, \bibinfo {author} {\bibfnamefont {C.~F.}\ \bibnamefont
  {Landes}}, \ and\ \bibinfo {author} {\bibfnamefont {S.}~\bibnamefont
  {Link}},\ }\bibfield  {title} {\enquote {\bibinfo {title} {Single-particle
  emission spectroscopy resolves d-hole relaxation in copper nanocubes},}\
  }\href {\doibase 10.1021/acsenergylett.9b01747} {\bibfield  {journal}
  {\bibinfo  {journal} {ACS Energy Letters}\ }\textbf {\bibinfo {volume} {4}},\
  \bibinfo {pages} {2458--2465} (\bibinfo {year} {2019})}\BibitemShut {NoStop}%
\bibitem [{\citenamefont {Zijlstra}, \citenamefont {Chon},\ and\ \citenamefont
  {Gu}(2009)}]{app1}%
  \BibitemOpen
  \bibfield  {author} {\bibinfo {author} {\bibfnamefont {P.}~\bibnamefont
  {Zijlstra}}, \bibinfo {author} {\bibfnamefont {J.~W.~M.}\ \bibnamefont
  {Chon}}, \ and\ \bibinfo {author} {\bibfnamefont {M.}~\bibnamefont {Gu}},\
  }\bibfield  {title} {\enquote {\bibinfo {title} {Five-dimensional optical
  recording mediated by surface plasmons in gold nanorods},}\ }\href {\doibase
  10.1038/nature08053} {\bibfield  {journal} {\bibinfo  {journal} {Nature}\
  }\textbf {\bibinfo {volume} {459}},\ \bibinfo {pages} {410--413} (\bibinfo
  {year} {2009})}\BibitemShut {NoStop}%
\bibitem [{\citenamefont {Taylor}, \citenamefont {Kim},\ and\ \citenamefont
  {Chon}(2012)}]{app2}%
  \BibitemOpen
  \bibfield  {author} {\bibinfo {author} {\bibfnamefont {A.~B.}\ \bibnamefont
  {Taylor}}, \bibinfo {author} {\bibfnamefont {J.}~\bibnamefont {Kim}}, \ and\
  \bibinfo {author} {\bibfnamefont {J.~W.~M.}\ \bibnamefont {Chon}},\
  }\bibfield  {title} {\enquote {\bibinfo {title} {Detuned surface plasmon
  resonance scattering of gold nanorods for continuous wave multilayered
  optical recording and readout},}\ }\href {\doibase 10.1364/OE.20.005069}
  {\bibfield  {journal} {\bibinfo  {journal} {Optics Express}\ }\textbf
  {\bibinfo {volume} {20}},\ \bibinfo {pages} {5069--5081} (\bibinfo {year}
  {2012})}\BibitemShut {NoStop}%
\bibitem [{\citenamefont {Lu}\ \emph {et~al.}(2012)\citenamefont {Lu},
  \citenamefont {Hou}, \citenamefont {Zhang}, \citenamefont {Liu},
  \citenamefont {Shen}, \citenamefont {Luo},\ and\ \citenamefont
  {Gong}}]{app3}%
  \BibitemOpen
  \bibfield  {author} {\bibinfo {author} {\bibfnamefont {G.}~\bibnamefont
  {Lu}}, \bibinfo {author} {\bibfnamefont {L.}~\bibnamefont {Hou}}, \bibinfo
  {author} {\bibfnamefont {T.}~\bibnamefont {Zhang}}, \bibinfo {author}
  {\bibfnamefont {J.}~\bibnamefont {Liu}}, \bibinfo {author} {\bibfnamefont
  {H.}~\bibnamefont {Shen}}, \bibinfo {author} {\bibfnamefont {C.}~\bibnamefont
  {Luo}}, \ and\ \bibinfo {author} {\bibfnamefont {Q.}~\bibnamefont {Gong}},\
  }\bibfield  {title} {\enquote {\bibinfo {title} {Plasmonic sensing via
  photoluminescence of individual gold nanorod},}\ }\href {\doibase
  10.1021/jp309450b} {\bibfield  {journal} {\bibinfo  {journal} {The Journal of
  Physical Chemistry C}\ }\textbf {\bibinfo {volume} {116}},\ \bibinfo {pages}
  {25509--25516} (\bibinfo {year} {2012})}\BibitemShut {NoStop}%
\bibitem [{\citenamefont {Zhang}\ \emph {et~al.}(2018)\citenamefont {Zhang},
  \citenamefont {Yu}, \citenamefont {Wei}, \citenamefont {Liu}, \citenamefont
  {Zhao},\ and\ \citenamefont {Huang}}]{app4}%
  \BibitemOpen
  \bibfield  {author} {\bibinfo {author} {\bibfnamefont {K.~Y.}\ \bibnamefont
  {Zhang}}, \bibinfo {author} {\bibfnamefont {Q.}~\bibnamefont {Yu}}, \bibinfo
  {author} {\bibfnamefont {H.}~\bibnamefont {Wei}}, \bibinfo {author}
  {\bibfnamefont {S.}~\bibnamefont {Liu}}, \bibinfo {author} {\bibfnamefont
  {Q.}~\bibnamefont {Zhao}}, \ and\ \bibinfo {author} {\bibfnamefont
  {W.}~\bibnamefont {Huang}},\ }\bibfield  {title} {\enquote {\bibinfo {title}
  {Long-lived emissive probes for time-resolved photoluminescence bioimaging
  and biosensing},}\ }\href {\doibase 10.1021/acs.chemrev.7b00425} {\bibfield
  {journal} {\bibinfo  {journal} {Chemical Reviews}\ }\textbf {\bibinfo
  {volume} {118}},\ \bibinfo {pages} {1770--1839} (\bibinfo {year}
  {2018})}\BibitemShut {NoStop}%
\bibitem [{\citenamefont {Zhang}\ \emph {et~al.}(2013)\citenamefont {Zhang},
  \citenamefont {Shen}, \citenamefont {Lu}, \citenamefont {Liu}, \citenamefont
  {He}, \citenamefont {Wang},\ and\ \citenamefont {Gong}}]{app5}%
  \BibitemOpen
  \bibfield  {author} {\bibinfo {author} {\bibfnamefont {T.}~\bibnamefont
  {Zhang}}, \bibinfo {author} {\bibfnamefont {H.}~\bibnamefont {Shen}},
  \bibinfo {author} {\bibfnamefont {G.}~\bibnamefont {Lu}}, \bibinfo {author}
  {\bibfnamefont {J.}~\bibnamefont {Liu}}, \bibinfo {author} {\bibfnamefont
  {Y.}~\bibnamefont {He}}, \bibinfo {author} {\bibfnamefont {Y.}~\bibnamefont
  {Wang}}, \ and\ \bibinfo {author} {\bibfnamefont {Q.}~\bibnamefont {Gong}},\
  }\bibfield  {title} {\enquote {\bibinfo {title} {Single bipyramid plasmonic
  antenna orientation determined by direct photoluminescence pattern
  imaging},}\ }\href {\doibase https://doi.org/10.1002/adom.201200041}
  {\bibfield  {journal} {\bibinfo  {journal} {Advanced Optical Materials}\
  }\textbf {\bibinfo {volume} {1}},\ \bibinfo {pages} {335--342} (\bibinfo
  {year} {2013})}\BibitemShut {NoStop}%
\bibitem [{\citenamefont {Lu}\ \emph {et~al.}(2015)\citenamefont {Lu},
  \citenamefont {Wang}, \citenamefont {Chou}, \citenamefont {Shen},
  \citenamefont {He}, \citenamefont {Cheng},\ and\ \citenamefont
  {Gong}}]{app6}%
  \BibitemOpen
  \bibfield  {author} {\bibinfo {author} {\bibfnamefont {G.}~\bibnamefont
  {Lu}}, \bibinfo {author} {\bibfnamefont {Y.}~\bibnamefont {Wang}}, \bibinfo
  {author} {\bibfnamefont {R.~Y.}\ \bibnamefont {Chou}}, \bibinfo {author}
  {\bibfnamefont {H.}~\bibnamefont {Shen}}, \bibinfo {author} {\bibfnamefont
  {Y.}~\bibnamefont {He}}, \bibinfo {author} {\bibfnamefont {Y.}~\bibnamefont
  {Cheng}}, \ and\ \bibinfo {author} {\bibfnamefont {Q.}~\bibnamefont {Gong}},\
  }\bibfield  {title} {\enquote {\bibinfo {title} {Directional side scattering
  of light by a single plasmonic trimer},}\ }\href {\doibase
  https://doi.org/10.1002/lpor.201500089} {\bibfield  {journal} {\bibinfo
  {journal} {Laser \& Photonics Reviews}\ }\textbf {\bibinfo {volume} {9}},\
  \bibinfo {pages} {530--537} (\bibinfo {year} {2015})}\BibitemShut {NoStop}%
\bibitem [{\citenamefont {Carattino}, \citenamefont {Caldarola},\ and\
  \citenamefont {Orrit}(2018)}]{app9}%
  \BibitemOpen
  \bibfield  {author} {\bibinfo {author} {\bibfnamefont {A.}~\bibnamefont
  {Carattino}}, \bibinfo {author} {\bibfnamefont {M.}~\bibnamefont
  {Caldarola}}, \ and\ \bibinfo {author} {\bibfnamefont {M.}~\bibnamefont
  {Orrit}},\ }\bibfield  {title} {\enquote {\bibinfo {title} {Gold
  nanoparticles as absolute nanothermometers},}\ }\href {\doibase
  10.1021/acs.nanolett.7b04145} {\bibfield  {journal} {\bibinfo  {journal}
  {Nano Letters}\ }\textbf {\bibinfo {volume} {18}},\ \bibinfo {pages}
  {874--880} (\bibinfo {year} {2018})}\BibitemShut {NoStop}%
\bibitem [{\citenamefont {Zhang}\ \emph {et~al.}(2019)\citenamefont {Zhang},
  \citenamefont {Han}, \citenamefont {Cao}, \citenamefont {Gao}, \citenamefont
  {Jia}, \citenamefont {Liu},\ and\ \citenamefont {Zeng}}]{app7}%
  \BibitemOpen
  \bibfield  {author} {\bibinfo {author} {\bibfnamefont {H.}~\bibnamefont
  {Zhang}}, \bibinfo {author} {\bibfnamefont {W.}~\bibnamefont {Han}}, \bibinfo
  {author} {\bibfnamefont {X.}~\bibnamefont {Cao}}, \bibinfo {author}
  {\bibfnamefont {T.}~\bibnamefont {Gao}}, \bibinfo {author} {\bibfnamefont
  {R.}~\bibnamefont {Jia}}, \bibinfo {author} {\bibfnamefont {M.}~\bibnamefont
  {Liu}}, \ and\ \bibinfo {author} {\bibfnamefont {W.}~\bibnamefont {Zeng}},\
  }\bibfield  {title} {\enquote {\bibinfo {title} {Gold nanoclusters as a
  near-infrared fluorometric nanothermometer for living cells},}\ }\href
  {\doibase 10.1007/s00604-019-3460-3} {\bibfield  {journal} {\bibinfo
  {journal} {Microchimica Acta}\ }\textbf {\bibinfo {volume} {186}},\ \bibinfo
  {pages} {353} (\bibinfo {year} {2019})}\BibitemShut {NoStop}%
\bibitem [{\citenamefont {Hastman}\ \emph {et~al.}(2020)\citenamefont
  {Hastman}, \citenamefont {Melinger}, \citenamefont {Aragonés}, \citenamefont
  {Cunningham}, \citenamefont {Chiriboga}, \citenamefont {Salvato},
  \citenamefont {Salvato}, \citenamefont {Brown}, \citenamefont {Mathur},
  \citenamefont {Medintz}, \citenamefont {Oh},\ and\ \citenamefont
  {Díaz}}]{app8}%
  \BibitemOpen
  \bibfield  {author} {\bibinfo {author} {\bibfnamefont {D.~A.}\ \bibnamefont
  {Hastman}}, \bibinfo {author} {\bibfnamefont {J.~S.}\ \bibnamefont
  {Melinger}}, \bibinfo {author} {\bibfnamefont {G.~L.}\ \bibnamefont
  {Aragonés}}, \bibinfo {author} {\bibfnamefont {P.~D.}\ \bibnamefont
  {Cunningham}}, \bibinfo {author} {\bibfnamefont {M.}~\bibnamefont
  {Chiriboga}}, \bibinfo {author} {\bibfnamefont {Z.~J.}\ \bibnamefont
  {Salvato}}, \bibinfo {author} {\bibfnamefont {T.~M.}\ \bibnamefont
  {Salvato}}, \bibinfo {author} {\bibfnamefont {C.~W.}\ \bibnamefont {Brown}},
  \bibinfo {author} {\bibfnamefont {D.}~\bibnamefont {Mathur}}, \bibinfo
  {author} {\bibfnamefont {I.~L.}\ \bibnamefont {Medintz}}, \bibinfo {author}
  {\bibfnamefont {E.}~\bibnamefont {Oh}}, \ and\ \bibinfo {author}
  {\bibfnamefont {S.~A.}\ \bibnamefont {Díaz}},\ }\bibfield  {title} {\enquote
  {\bibinfo {title} {Femtosecond laser pulse excitation of dna-labeled gold
  nanoparticles: Establishing a quantitative local nanothermometer for
  biological applications},}\ }\href {\doibase 10.1021/acsnano.0c02899}
  {\bibfield  {journal} {\bibinfo  {journal} {ACS Nano}\ }\textbf {\bibinfo
  {volume} {14}},\ \bibinfo {pages} {8570--8583} (\bibinfo {year}
  {2020})}\BibitemShut {NoStop}%
\bibitem [{\citenamefont {Shahbazyan}(2013)}]{mech1}%
  \BibitemOpen
  \bibfield  {author} {\bibinfo {author} {\bibfnamefont {T.~V.}\ \bibnamefont
  {Shahbazyan}},\ }\bibfield  {title} {\enquote {\bibinfo {title} {Theory of
  plasmon-enhanced metal photoluminescence},}\ }\href {\doibase
  10.1021/nl303851z} {\bibfield  {journal} {\bibinfo  {journal} {Nano Letters}\
  }\textbf {\bibinfo {volume} {13}},\ \bibinfo {pages} {194--198} (\bibinfo
  {year} {2013})}\BibitemShut {NoStop}%
\bibitem [{\citenamefont {Dulkeith}\ \emph {et~al.}(2004)\citenamefont
  {Dulkeith}, \citenamefont {Niedereichholz}, \citenamefont {Klar},
  \citenamefont {Feldmann}, \citenamefont {von Plessen}, \citenamefont
  {Gittins}, \citenamefont {Mayya},\ and\ \citenamefont {Caruso}}]{mech2}%
  \BibitemOpen
  \bibfield  {author} {\bibinfo {author} {\bibfnamefont {E.}~\bibnamefont
  {Dulkeith}}, \bibinfo {author} {\bibfnamefont {T.}~\bibnamefont
  {Niedereichholz}}, \bibinfo {author} {\bibfnamefont {T.~A.}\ \bibnamefont
  {Klar}}, \bibinfo {author} {\bibfnamefont {J.}~\bibnamefont {Feldmann}},
  \bibinfo {author} {\bibfnamefont {G.}~\bibnamefont {von Plessen}}, \bibinfo
  {author} {\bibfnamefont {D.~I.}\ \bibnamefont {Gittins}}, \bibinfo {author}
  {\bibfnamefont {K.~S.}\ \bibnamefont {Mayya}}, \ and\ \bibinfo {author}
  {\bibfnamefont {F.}~\bibnamefont {Caruso}},\ }\bibfield  {title} {\enquote
  {\bibinfo {title} {Plasmon emission in photoexcited gold nanoparticles},}\
  }\href {\doibase 10.1103/PhysRevB.70.205424} {\bibfield  {journal} {\bibinfo
  {journal} {Physical Review B}\ }\textbf {\bibinfo {volume} {70}},\ \bibinfo
  {pages} {205424} (\bibinfo {year} {2004})}\BibitemShut {NoStop}%
\bibitem [{\citenamefont {Cheng}\ \emph {et~al.}(2018)\citenamefont {Cheng},
  \citenamefont {Zhang}, \citenamefont {Zhao}, \citenamefont {Wen},
  \citenamefont {Hu}, \citenamefont {Gong},\ and\ \citenamefont {Lu}}]{PL0}%
  \BibitemOpen
  \bibfield  {author} {\bibinfo {author} {\bibfnamefont {Y.}~\bibnamefont
  {Cheng}}, \bibinfo {author} {\bibfnamefont {W.}~\bibnamefont {Zhang}},
  \bibinfo {author} {\bibfnamefont {J.}~\bibnamefont {Zhao}}, \bibinfo {author}
  {\bibfnamefont {T.}~\bibnamefont {Wen}}, \bibinfo {author} {\bibfnamefont
  {A.}~\bibnamefont {Hu}}, \bibinfo {author} {\bibfnamefont {Q.}~\bibnamefont
  {Gong}}, \ and\ \bibinfo {author} {\bibfnamefont {G.}~\bibnamefont {Lu}},\
  }\bibfield  {title} {\enquote {\bibinfo {title} {Understanding
  photoluminescence of metal nanostructures based on an oscillator model},}\
  }\href {\doibase 10.1088/1361-6528/aac44f} {\bibfield  {journal} {\bibinfo
  {journal} {Nanotechnology}\ }\textbf {\bibinfo {volume} {29}},\ \bibinfo
  {pages} {315201} (\bibinfo {year} {2018})}\BibitemShut {NoStop}%
\bibitem [{\citenamefont {Zhang}\ \emph {et~al.}(2020)\citenamefont {Zhang},
  \citenamefont {Wen}, \citenamefont {Ye}, \citenamefont {Lin}, \citenamefont
  {Gong},\ and\ \citenamefont {Lu}}]{mech3}%
  \BibitemOpen
  \bibfield  {author} {\bibinfo {author} {\bibfnamefont {W.}~\bibnamefont
  {Zhang}}, \bibinfo {author} {\bibfnamefont {T.}~\bibnamefont {Wen}}, \bibinfo
  {author} {\bibfnamefont {L.}~\bibnamefont {Ye}}, \bibinfo {author}
  {\bibfnamefont {H.}~\bibnamefont {Lin}}, \bibinfo {author} {\bibfnamefont
  {Q.}~\bibnamefont {Gong}}, \ and\ \bibinfo {author} {\bibfnamefont
  {G.}~\bibnamefont {Lu}},\ }\bibfield  {title} {\enquote {\bibinfo {title}
  {Influence of non-equilibrium electron dynamics on photoluminescence of
  metallic nanostructures},}\ }\href {\doibase 10.1088/1361-6528/abb1ee}
  {\bibfield  {journal} {\bibinfo  {journal} {Nanotechnology}\ }\textbf
  {\bibinfo {volume} {31}},\ \bibinfo {pages} {495204} (\bibinfo {year}
  {2020})}\BibitemShut {NoStop}%
\bibitem [{\citenamefont {Prodan}\ \emph {et~al.}(2003)\citenamefont {Prodan},
  \citenamefont {Radloff}, \citenamefont {Halas},\ and\ \citenamefont
  {Nordlander}}]{cpm1}%
  \BibitemOpen
  \bibfield  {author} {\bibinfo {author} {\bibfnamefont {E.}~\bibnamefont
  {Prodan}}, \bibinfo {author} {\bibfnamefont {C.}~\bibnamefont {Radloff}},
  \bibinfo {author} {\bibfnamefont {N.~J.}\ \bibnamefont {Halas}}, \ and\
  \bibinfo {author} {\bibfnamefont {P.}~\bibnamefont {Nordlander}},\ }\bibfield
   {title} {\enquote {\bibinfo {title} {A hybridization model for the plasmon
  response of complex nanostructures},}\ }\href {\doibase
  10.1126/science.1089171} {\bibfield  {journal} {\bibinfo  {journal}
  {Science}\ }\textbf {\bibinfo {volume} {302}},\ \bibinfo {pages} {419--422}
  (\bibinfo {year} {2003})}\BibitemShut {NoStop}%
\bibitem [{\citenamefont {Jain}, \citenamefont {Huang},\ and\ \citenamefont
  {El-Sayed}(2007)}]{cpm2}%
  \BibitemOpen
  \bibfield  {author} {\bibinfo {author} {\bibfnamefont {P.~K.}\ \bibnamefont
  {Jain}}, \bibinfo {author} {\bibfnamefont {W.}~\bibnamefont {Huang}}, \ and\
  \bibinfo {author} {\bibfnamefont {M.~A.}\ \bibnamefont {El-Sayed}},\
  }\bibfield  {title} {\enquote {\bibinfo {title} {On the universal scaling
  behavior of the distance decay of plasmon coupling in metal nanoparticle
  pairs: A plasmon ruler equation},}\ }\href {\doibase 10.1021/nl071008a}
  {\bibfield  {journal} {\bibinfo  {journal} {Nano Letters}\ }\textbf {\bibinfo
  {volume} {7}},\ \bibinfo {pages} {2080--2088} (\bibinfo {year}
  {2007})}\BibitemShut {NoStop}%
\bibitem [{\citenamefont {Griffiths}(2013)}]{Griffiths}%
  \BibitemOpen
  \bibfield  {author} {\bibinfo {author} {\bibfnamefont {D.~J.}\ \bibnamefont
  {Griffiths}},\ }\href@noop {} {\emph {\bibinfo {title} {Introduction to
  Electrodynamics (4rd Edition)}}}\ (\bibinfo  {publisher} {Pearson},\ \bibinfo
  {year} {2013})\BibitemShut {NoStop}%
\bibitem [{\citenamefont {Su}\ \emph {et~al.}(2003)\citenamefont {Su},
  \citenamefont {Wei}, \citenamefont {Zhang}, \citenamefont {Mock},
  \citenamefont {Smith},\ and\ \citenamefont {Schultz}}]{cp1}%
  \BibitemOpen
  \bibfield  {author} {\bibinfo {author} {\bibfnamefont {K.-H.}\ \bibnamefont
  {Su}}, \bibinfo {author} {\bibfnamefont {Q.-H.}\ \bibnamefont {Wei}},
  \bibinfo {author} {\bibfnamefont {X.}~\bibnamefont {Zhang}}, \bibinfo
  {author} {\bibfnamefont {J.~J.}\ \bibnamefont {Mock}}, \bibinfo {author}
  {\bibfnamefont {D.~R.}\ \bibnamefont {Smith}}, \ and\ \bibinfo {author}
  {\bibfnamefont {S.}~\bibnamefont {Schultz}},\ }\bibfield  {title} {\enquote
  {\bibinfo {title} {Interparticle coupling effects on plasmon resonances of
  nanogold particles},}\ }\href {\doibase 10.1021/nl034197f} {\bibfield
  {journal} {\bibinfo  {journal} {Nano Letters}\ }\textbf {\bibinfo {volume}
  {3}},\ \bibinfo {pages} {1087--1090} (\bibinfo {year} {2003})}\BibitemShut
  {NoStop}%
\bibitem [{\citenamefont {S{\"o}nnichsen}\ \emph {et~al.}(2005)\citenamefont
  {S{\"o}nnichsen}, \citenamefont {Reinhard}, \citenamefont {Liphardt},\ and\
  \citenamefont {Alivisatos}}]{cp2}%
  \BibitemOpen
  \bibfield  {author} {\bibinfo {author} {\bibfnamefont {C.}~\bibnamefont
  {S{\"o}nnichsen}}, \bibinfo {author} {\bibfnamefont {B.~M.}\ \bibnamefont
  {Reinhard}}, \bibinfo {author} {\bibfnamefont {J.}~\bibnamefont {Liphardt}},
  \ and\ \bibinfo {author} {\bibfnamefont {A.~P.}\ \bibnamefont {Alivisatos}},\
  }\bibfield  {title} {\enquote {\bibinfo {title} {A molecular ruler based on
  plasmon coupling of single gold and silver nanoparticles},}\ }\href {\doibase
  10.1038/nbt1100} {\bibfield  {journal} {\bibinfo  {journal} {Nature
  Biotechnology}\ }\textbf {\bibinfo {volume} {23}},\ \bibinfo {pages}
  {741--745} (\bibinfo {year} {2005})}\BibitemShut {NoStop}%
\bibitem [{\citenamefont {Hu}\ \emph {et~al.}(2012)\citenamefont {Hu},
  \citenamefont {Duan}, \citenamefont {Yang},\ and\ \citenamefont
  {Shen}}]{cp3}%
  \BibitemOpen
  \bibfield  {author} {\bibinfo {author} {\bibfnamefont {H.}~\bibnamefont
  {Hu}}, \bibinfo {author} {\bibfnamefont {H.}~\bibnamefont {Duan}}, \bibinfo
  {author} {\bibfnamefont {J.~K.~W.}\ \bibnamefont {Yang}}, \ and\ \bibinfo
  {author} {\bibfnamefont {Z.~X.}\ \bibnamefont {Shen}},\ }\bibfield  {title}
  {\enquote {\bibinfo {title} {Plasmon-modulated photoluminescence of
  individual gold nanostructures},}\ }\href {\doibase 10.1021/nn3039066}
  {\bibfield  {journal} {\bibinfo  {journal} {ACS Nano}\ }\textbf {\bibinfo
  {volume} {6}},\ \bibinfo {pages} {10147--10155} (\bibinfo {year}
  {2012})}\BibitemShut {NoStop}%
\end{thebibliography}%

\end{document}